# Platoon–assisted Vehicular Cloud in VANET: Vision and Challenges

Meysam Nasimi[1], Mohammad Asif Habibi[1], and Hans D. Schotten[1,2]

[1] Institute of Wireless Communication, Technische Universität Kaiserslautern, 67663 Kaiserslautern, Germany
[2] Research Group Intelligent Networks, German Research Center for Artificial Intelligence (DFKI GmbH), Kaiserslautern, Country
Email: {nasimi, asif, schotten}@eit.uni-kl.de; hans_dieter.schotten@dfki.de

*Abstract* —Intelligent connected vehicles equipped with wireless sensors, intelligent control system, and communication devices are expected to commercially launch and emerge on road in short-term. These smart vehicles are able to partially/fully drive themselves; collect data from sensors, make and execute decisions based on that data; communicate with other vehicles, pedestrians, and nodes installed on the road; and provide infotainment and value-added services, such as broadband transmission of ultra-high definition video, files/apps downloading and uploading, online gaming, access to social media, audio/video conference streaming (office-in-car), live TV streaming, etc.; and so on. In addition, it is also possible for autonomous vehicles to form a "platoon" on road; maintaining close proximity in order to reduce the consumption of fuel and/or emission of gas, decrease costs, increase safety, and enhance the efficiency of the legacy transportation system. These emerging vehicular applications demand a large amount of computing and communication capacity to excel in their compute-intensive and latency-sensitive tasks. Based on these facts, the authors of this paper presented a visionary concept – "platoon-assisted vehicular cloud" – that exploits underutilized resources in platoons to augment vehicular cloud aiming to provide cost-effective and on-demand computing resources. Moreover, the authors presented three potential scenarios and explained the exploitation of platoon resources and roadside infrastructure to facilitate new applications. Besides system design, the paper did also summarize a number of open research challenges with the purpose of motivating new advances and potential solutions to this field.

*Index Terms*—Cloud computing, computing-on-wheels, mobile edge computing, vehicular cloud computing, vehicle as a resource

## I. Introduction

Intelligent transportation system (ITS) provides efficient solutions to improve the performance of legacy transportation systems by applying the most advanced information and communication technologies, and the most recent control and sensor systems, such as cooperative communication on the road and/or with the pedestrian, automotive wireless sensors and intelligent control to vehicles. These technologies are applied to smart vehicles and digitized-transportation infrastructure in order to collect real-time data for road users and transportation operators. The data is then used by both smart vehicles and transportation operators to make and execute better decisions, which leads to reduce traffic congestion, increase road capacity, enhance travel safety, reduce atmospheric emission, and improve homeland security [1]. To achieve these goals and unlock new business opportunities in transportation sector, we first and foremost need to architect and apply a communication infrastructure that is suitable to fulfill its operation, functional, and performance requirements.

In order to establish communication links between vehicles, vehicles and transportation operators, and vehicles and networks in ITS; a reliable and efficient communication system is crucial to design. A promising communication concept that has attracted tremendous interest and extensive research and development activities from car manufacturing industries, transportation authorities (governments), research community, and standards organizations is the vehicular ad-hoc network (VANET). It is one of the applications conceived in ITS that allows participating vehicles to form a self-organized network without the need for a fixed physical communication infrastructure [2], [3].

With the advancement of smart vehicles in VANET, an increasing number of interactive applications and services have been emerged and are currently being used. Many of these applications, such as autonomous driving, image processing, and three-dimensional (3D) navigation, are computational-intensive, delay-sensitive, and bandwidth-demanding. However, due to the limited computing and storage resources provided by the vehicles' onboard units (OBUs), it is difficult to satisfy the requirements of these applications. To overcome these issues, task offloading to remote centralized cloud (CC) [4] has been regarded as a promising solution. The vehicles have an opportunity to offload their complex computational tasks to the centralized cloud server, using abundant cloud resources to help them process the tasks. This offloading scheme effectively reduces computational load of vehicles, as the quality-of-service (QoS) requirements increase, its defects are becoming more and more obvious.

Offloading computation tasks to the centralized server, which are often far away from the service requester, lead to long latency in the transmission process of the task.

This work was supported by the European Union Horizon-2020 Projects 5G-AuRA under Grant 675806.

Corresponding author email: nasimi@eit.un-kl.de.

This means that all the data and requests need to traverse the backhaul and backbone networks (e.g., base stations, routers, etc.). However, due to real-time requirements of the data transmission and service control of VANET, the data analysis and control logic in the remote cloud center cannot guarantee the delay and jitter performance of the applications. In addition, rapid increase in the number of users and service types give rise to drastic increase in accessing the remote cloud, which result in excessive network load. This leads to problems such as increased delay and severe congestion, which seriously degrade offloading performance. Therefore, the computation offloading method of CC is not suitable and efficient for the delay-sensitive task in VANET.

In order to meet the stringent requirements of delay-sensitive tasks, the concept of vehicular edge computing (VEC), which is the integration of mobile edge computing (MEC) in VANET has been proposed. In this approach, MEC provides ultra reliable, high bandwidth, and low-latency computing services for vehicles by extending cloud computing to the edge of the network. This is because the computing resources are in the proximity of the vehicles, and the vehicles usually needs only a single hop to access the edge cloud, which greatly reduces the communication delay. Moreover, the MEC server at the edge can optimize the service dynamically. This can be done with auxiliary information obtained from the environmental context in real time.

With the increasing demand for delay sensitive applications, MEC has been used in many scenarios, such as augmented reality and virtual reality. In VANET, a significant number of researches have been carried out on the applications of MEC. In the existing studies, MEC servers are usually deployed near RSU, so their performance will be limited by the infrastructure. To be more precise: First, vehicles need to pay for the transmission of their computation's tasks since the spectrum cost of infrastructure is expensive. Second, due to the limited communication range and sparse deployment, the infrastructure cannot completely cover vehicles traveling on the road. Third, in terms of service reliability, especially when the infrastructure is damaged or user requests exceed its processing capacity, service cannot be provided.

In order to support VEC in VANET, the promising concept of cooperative vehicular cloud computing (CVCC) was introduced. CVCC is a group of vehicles with computing capabilities, which facilitate computation-intensive task by sharing their surplus computation and storage resources [5]. Different from the centralized or edge cloud, CVCC has its unique features, which offer computation service without need for a fixed infrastructure. However, one main concern regarding this scheme is how to encourage vehicles to participate in cooperative task offloading paradigm.

Platooning plays assistance role in this direction by integration of potential underutilized resources, offered by platoons into the vehicular cloud eco-system. In this way, we can enhance the scalability of computing services for vehicles, and reduce the cost of using cloud resources. Platooning is basically one of the key technologies of autonomous driving in which a fleet of vehicles – composed of a head vehicle "leader" and a number of following vehicles "followers" – with only a few meters between them [6]. This feature has brought a number of advantages to the transportation industry including significant safety, energy efficiency, and cost benefits.

To elaborate a little further, the platoons are stimulated by the autonomous vehicles (e.g., cars, trucks, vans, etc.) and the vehicular resources that are readily available. Specially, truck platooning plays a key role on the way towards the future road transport system. To better perceive the truck platoons' potentialities, by 2025 around 70% of all freight transport will travel on public roads, making truck platoon as a key enabler for the revolutionizing of transportation system. Besides, major truck companies (such as DAF, Daimler, Iveco, MAN, Scania, Volvo, and tesla) as well as other key players (such as amazon and DHL) are at the forefront to intensively have a look at this innovative technology. Moreover, the government agencies such as the U.S. Department of Transportation, national governments and the European Union (EU) are embracing new autonomous vehicle technologies including the platooning systems. As of now, 16 U.S. states support demonstrations and testing, and 11 states have laws regarding truck platooning, according to the Federal Highway Administration. In Europe, appropriate regulatory framework is created at both EU and international levels, building on the 2016 Declaration of Amsterdam and adapting existing United Nations (UN), EU, and national legislations. All in all, the outlook for platooning is bright, so much so that the massive numbers of truck platoon as underutilized cloud resources is able to augment the vehicular cloud [7], [8], [9].

Based on these facts, we propose a novel concept called "platoon-assisted vehicular cloud" towards the integration of platooning into the vehicular cloud eco-system in this article. We investigate three scenarios for platooning, showing how the proposed approaches support applications' computations needs. These scenarios are: i) collaborative inter-platoon task offloading, ii) platoon-assisted vehicular for vehicles, and iii) support for MEC.

The remainder of this paper is organized as follows: we commence this article by providing key backgrounds behind ITS, platooning in autonomous driving, and VANET in Sec. II. Following that, we provide a detailed discussion and systems design of proposed concept and its three scenarios in Sec. III. In addition, we discuss challenges and open research issues related to the proposed concept in Sec. V. Finally, we conclude the paper with Sec. VI.

## II. MOTIVATION

In this section, we provide motivation and key backgrounds related to our proposed concept and three scenarios for platoon-assisted computation in vehicular cloud. Our proposal is motivated by recent advances in ITS, particularly in automated driving vehicles, the emerging low-latency, high bandwidth vehicle to vehicle (V2V) and vehicle to infrastructure (V2I) communication technologies, and edge computing technologies. Therefore, it is essential to provide some fundamentals about these topics in this section, before we move to the next section where the core idea of this research is presented.

### A. Intelligent Transportation System

ITS is an efficient transportation solution that enhances performance, increases road capacity, improves travel security, and reduces atmospheric emission and traffic congestion in legacy transportation system. With the deployment of ITS, drivers, roadside infrastructure and road authorities, are provided with accurate information about the status, intentions, conditions, and early warnings of vehicles [10]. Moreover, nearby inter-connected vehicles also exchange information with one another. This leads to improve traffic safety and enhance efficiency of legacy transportation system. To accomplish these goals, ITS integrates and applies the most advanced information and communication technologies, and the most recent intelligent control and wireless sensor systems.

Over the last two decades, ITS has been widely applied for various purposes. However, the applications of ITS are not just limited to the interconnection between vehicles and collection of information, but also to provide innovative services in order to make transport system safer and smarter, more coordinated, and cost/energy efficient. Some recent attractive applications of ITS are: i) emergency vehicle notification systems, ii) automatic road enforcement, iii) variable speed limits, iv) dynamic traffic light sequence, v) collision avoidance systems, vi) night vision enhancement, vii) intelligent cruise control and lane keeping assistance [10], [11].

In addition to above-mentioned features and applications, ITS has a wide set of applications. All these applications are put into three categories; namely, advanced driver assistance systems (ADAS), advanced traveler information systems (ATIS), and advanced traffic management systems (ATMS) [10]. These set of applications are discussed as follows:
- ADAS includes applications such as cooperative and intersection collision warnings, slow indications of vehicles, lane change messages, speed control of vehicles, and reverse parking assistance for vehicles.
- ATIS includes applications such as information about public transport, trip reservation facility for passengers, route planning service, online booking of seats, local electronic commerce, and trip matching services.
- ATMS includes applications such as dynamic route information for vehicles, dynamic lane assignment for smart vehicles, hazardous location that should be detected, deterioration detection, and incident detection.

ITS is expected to cope with future demands for system, service, and business beyond the horizon of 2020, play a vital role in enabling of new innovations, and prove a significant benefit to the economic output. Many countries and regions, such as EU, US, Japan, UK, South Korea, etc., are actively participating the race towards ITS, attempting to establish their technical and economic leaderships in the next decade. For example: Germany has recently opened the first-ever electric-Highway (eHighway) system in the world, which ensures an extremely reliable energy supply for truck platooning in heavily-used truck routes [12].

### B. Platooning in Autonomous Driving

Autonomous driving is a promising technology that is currently receiving so much attention from car manufacturers, communication industry, transportation authorities and academia. It encompasses of a number of technologies (e.g., wireless communication, control and autonomous), various types of infrastructures (e.g., transportation, communication and safety/security), differentiated capabilities and contexts (i.e., the surround view, traffic sign recognition and self-emergency braking), varieties of use cases and business cases, and different products and services. In autonomous driving system, smart vehicles collect information, makes decision according to the collected information, and execute the decision. This information comes from sensors that are installed on vehicles, and physical and digital infrastructures (both communication and transportation). Many of these technologies are capable of guiding vehicles and/or drive vehicles with minimal or no driver input (especially in test situations or across diverse driving environments). These technologies added various new applications, features, and services to smart vehicle. One of its main features is platooning.

Platooning of smart vehicles is basically an innovative and automated way of driving of a cluster of smart vehicles that aims at reducing fuel consumption and gas emission, achieving safe and efficient transport [13]. In platooning, a group of vehicles comprised of a head vehicle (leader) and several following vehicles called platoon-members (or followers) are traveling on the highway in a close proximity to one another. Platoon-members must act cooperatively to control and manage the platoon actions including platoon formation, merging, splitting, maintenance, etc. The operation of these actions is facilitated by the improved version of adaptive cruise control, which assists the platoons in matching the movement of a vehicle to the distance, speed, and direction of the vehicle in front [14]–[16]. It is worth mentioning that there are two possible ways to form platoons, the normal platoons and the customized truck platoons, which are collectively referred as platoons. The former occurs in regular highways, in which different types of vehicles (e.g., cars, vans, trucks, buses, etc.) form a platoon. The latter takes place in smart highways such as eHighways (i.e., eHighway is a highway equipped with overhead cables to charge the trucks as they drive), in which the trucks are able to connect to the overhead electric cables (i.e., inspired by electric-train lines). A key difference is in the latter case, where the chance of more trucks involve in platooning is increased as they can reduce their travel cost by taking advantage of the overhead electricity cable.

The idea of platooning was originally proposed during the eighties in the partners for advanced transit and highways (PATH) project at the University of California Berkeley [17]. The objective of the project was to increase highway capacity with minimum changes to existing infrastructure. In Europe, several projects on platooning have been carried out, such as safe road trains for the environment (SARTE) [18], cooperative dynamic formation of platoons for safe and energy-optimized goods transportation (COMPANION) [19], enabling safe multi-brand platooning for Europe (Ensemble) [20], European truck platooning challenge (ETPC) [21], and a recent truck platooning on electric-highway (eHighway) near to Frankfurt in Germany [12]. The SARTRE project is funded by the European Commission with the aim to increase fuel efficiency, reduce congestion, and increase safety. COMPANION aims at the dynamic forming of platoons and is supported by Volkswagen and Scania. The ETPC is being organized by the Netherlands to promote platooning by bringing truck convoys to public roads. The goal of the ENSEMBLE project is to accelerate the adoption of multi-brand truck platooning. Most importantly, all major truck manufacturers have developed technologies that allow platooning, and several field tests are planned or are currently taking place.

The major players of platooning can be truck and automobile manufacturers (e.g. Volvo, BMW, etc.), logistics companies (DHL, DB Schenker, etc.), edge computing platform providers (e.g. IBM, HPE, Cisco, etc.), vehicular application providers, virtualization technologies (e.g. NFV/VNF), mobile network operators (e.g. Vodafone, AT&T), and public transport authority.

Platooning offers various advantages. First, it has the potential to lowers fuel consumption and environmental pollution. Second, it improves traffic flow and reduces jams on highways [16], [22]. Third, platooning enhances traffic safety as the likelihood of accidents due to human error is reduced [23]. Last but not least, due to the relatively fixed position of vehicles within the same platoon, the cooperative communication and computation application can be facilitated. Consequently, vehicular networking and computing can be improved. Therefore, the stability is increased dramatically in platooning.

Despite above key advantages of platooning, there are still a number of challenges that remain to be tackled. The first challenge is rooted in multidisciplinary nature of platooning in which several research fields are involves, such as control theory, communication, traffic engineering. For example: platoon stability, transportation network design, communication networks, and incentive design are the areas covered by platooning. The second challenge is to autonomously implement platooning in highways. Many factors may affect the incentive to form platoons, such as transportation network design, different destinations for each vehicle, etc.

*C. Vehicular Ad-Hoc Network*

VANET is one of the major components of ITS that provides communication exchange among nearby vehicles and between vehicles and nearby fixed roadside equipment. It primarily concentrates on specifying communication protocols for V2V and V2I information exchange methods to enhance safe and cooperative driving, and traffic management. In VANET, the main communication modes can be among vehicles, and vehicles and road-side unit (RSU). Every vehicle is equipped with special electronic devices known as on-board unit (OBU) to facilitate computation and communication.

V2V as one of the communication modes of VANET is a direct communication link that occurs over sidelink air-interface (PC5) between nearby vehicles on a highway/road [24]. As the second type of communication mode of VANET, V2I is referred to a communication link that occurs over PC5 between vehicles and RSUs that are installed alongside the road. Both types of communication are wireless, and have the capabilities for transportation systems within a 1000 meter range at typical highway speeds.

Recently, there has been a number of significant enhancements in ITS and more specifically in VANET. One of these progresses was the deployment of advanced driver assistance system that was proposed to accelerate platooning process. It is obvious that platooning requires coordination among platoon-members to maintain closer possible headway between vehicles. However, this coordination cannot be handled efficiently by standard sensor-based adaptive cruise control (ACC) techniques [25]. Consequently, cooperative ACC (CACC) as an enhanced version of ACC can be applied to handle the coordination among the platoon-members. In CACC, vehicles follow each other more accurately, safely, and closely. Moreover, braking and accelerating are also done cooperatively and synchronously. Therefore, this new enhancement in VANET leads to accelerate the platooning. Indeed, CACC integrates V2V communication with traditional ACC to provide sufficient information (e.g., location, speed acceleration, braking information, etc.) about the vehicle it is following.

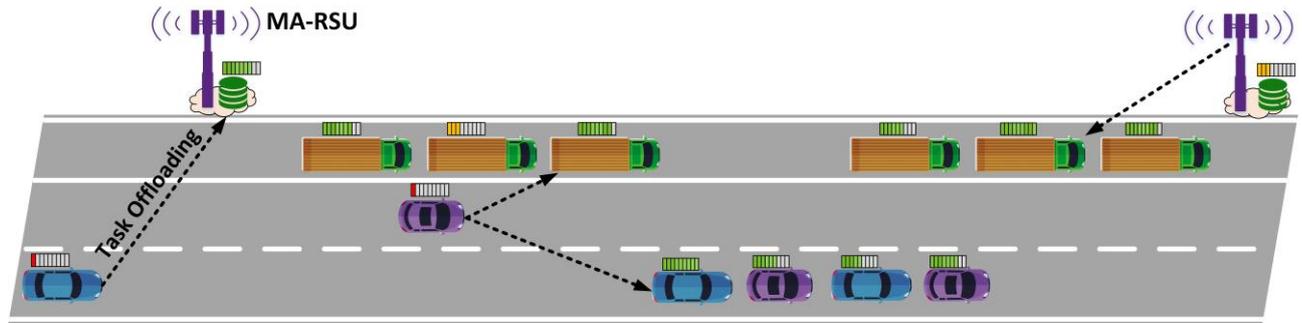

Fig. 1. System model of the future platooning.

Thanks to the advancement in VANET, CACC techniques can leverage newer protocols such as dedicated short-range communication (DSRC), wireless access in vehicular environments (WAVE), and communications access for land mobiles (CALM) [26]. CACC vehicles utilize beaconing to obtain information form neighborhood or infrastructure elements. Three of them are wireless communication protocols and air interfaces that are used for a variety of vehicular communication scenarios spanning multiple modes of communications and multiple methods of transmissions in intelligent transportation. Beacon messages usually contain the vehicle's identity, velocity, geographical location, and acceleration information. Thus, the beaconing process enables vehicles with environmental awareness information about neighborhood elements [27].

To be more precise, DSRC refers to a suite of standards that includes IEEE 802.11p and the IEEE 1609 stack. The former provides the medium access control (MAC) and physical (PHY) layers for communications in a vehicular environment, and the latter extend the IEEE 802.11p MAC layer functions for multi-channel operation as well as the specification of the upper layers. In addition to the technical work, European Telecommunication Standard Institute (ETSI) specified the first set of ITS-G5 communication protocols and architecture regulating operation in the 5 GHz spectrum for C-ITS. ITS-G5 reuses the PHY and MAC layers of the IEEE 802.11p framework [28].

## III. PROPOSED PLATOON-ASSISTED VEHICULAR CLOUD

In Fig. 1, the proposed platoon-assisted architecture for vehicular cloud is illustrated. The architecture is composed of the following two interacting components: the platoon cloud and MEC-assisted roadside unit (MA-RSU).

**Platoon cloud**: A cloud established among a group of vehicles in the form of a platoon. The vehicles in the platoon are viewed as mobile cloud sites that cooperatively share their computation resources. An inter-vehicle network is formed by V2V communication. Different from CVCC, in the platoon cloud, the cooperation of platoon-members provides stable cloud service in the long run which leads to higher computation service efficiency. There are three reasons for that. The first reason is the monetary incentive of vehicles to join the platoon i.e., joining a platoon will result in travel cost reduction. Second, the creation of platoons can be scheduled in advance based on the trip information of vehicles (e.g., origin location, a destination location, an earliest departure time, and a latest arrival time at the destination). Third, the vehicles in the platoon have longer encounter time and higher contact probability with each other. As a result, cooperation between platoon-members extends the computation capability of vehicular cloud by enhancing the overall resource utilization in VANET.

**MA-RSU:** MA-RSU is a small-scale site that offers cloud services to bypassing vehicles. For this purpose, the dedicated MEC server is attached to RSU which provides cloud service for vehicles in the communication range. The vehicles access MA-RSU by V2I communications. The MEC servers virtualize physical resources and act as a potential cloud site. RSUs are deployed along the roadside providing radio interfaces for vehicles to access the cloud. MA-RSU is accessible only to the vehicles within the radio coverage area of the MA-RSU. As the vehicle moves along the road, vehicles may pass through several MA-RSU during the task offloading process. Under such conditions, for service continuity, the customized virtual machine (VM) needs to be transferred across the MA-RSUs. This process is referred to as VM migration, in which VM should be synchronously transferred between involved MA-RSUs [29].

The proposed architecture has three key advantages: First, the physical resources on the road are fully utilized. From vehicles to roadside infrastructures, the computation resources are all merged into the cloud. All clouds are accessible to all vehicles. Second, the hierarchical nature of the architecture allows vehicles using different communication technologies to access different layers of clouds accordingly. Hence, the architecture is flexible and compatible with heterogeneous wireless technologies such as DSRC, long term evolution (LTE), LTE-advanced (LTE-A), and fifth generation (5G) of cellular network. Third, the platoon cloud and MA-RSUs are small-scale localized clouds. Such distributed clouds can be rapidly deployed and quickly provide services.

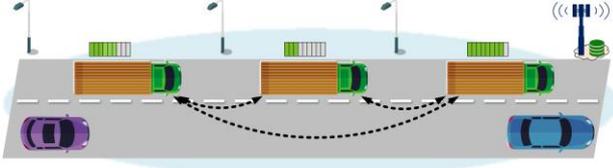

Fig. 2. Scenario of task offloading between platoon-members.

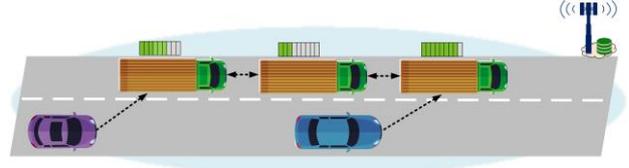

Fig. 3. Scenario of vehicles offload their tasks to platoons.

## IV. PROMISING APPLICATIONS OF PLATOON-ASSISTED VEHICULAR CLOUD

The proposed platoon-assisted vehicular cloud extends computation capability of vehicular cloud aiming to enhance overall resource utilization in VANET. In this section, we illustrate three potential application scenarios and explain the exploitation of a platoon cloud and MA-RSU to facilitate smart vehicles' intensive computational processing.

### A. Collaborative Inter-Platoon Task Offloading

Emerging vehicular applications, such as autonomous-driving, crowd-sensing, and voice processing are typically computing resource-hungry services; which require high-density computing resources. Therefore, the limited size of available resources in the vehicle cannot meet the service requirements, which in turn cause high application processing delay, and even lead to road congestion and accident. In this context, we propose the cooperative task offloading scheme for the vehicles within the same platoon, namely, inter-platoon task offloading.

In Fig. 2, vehicles in platoons are classified into two categories: First category includes the vehicles, which generate and offload computation task for cloud execution. Second category includes the vehicles with sufficient computing resources, which provide computing services. Note that the role of each vehicle in platoon is not fixed, and it can be changed depending on the sufficiency of its computing resources during the trip. In this scenario, the task offloading procedure between platoon-members within same platoon are as follow: First, the member that needs cloud service submits its request to the adjacent vehicles. Then, the members that receive the request, based on their computing resources sufficiency, inform the requester. After selecting the suitable member(s), the task will be uploaded for the processing. Finally, upon the completion of tasks execution, the result will be sent back.

In order to achieve efficient inter-platoon task offloading, there are still a number of technical challenges exist that need further investigation. The first challenge is to explore an efficient task scheduling and resource allocation scheme to offload tasks between platoon-members. The second challenge is to find an optimal task allocation policy by considering application latency constraint and computation load. Third challenge is intra-platoon task migration (e.g., offload task to another platoon in proximity, CVCC or MEC), this may be the case for a platoon with overloaded computational resources.

It is noteworthy that the work [30] considers inter-platoon task offloading with an objective to minimize the offloading cost within platoon within delay constraint. By adopting classical Lagrangian relaxation-based aggregated cost algorithm, they demonstrate that the decision of the task offloading can be well applied to the platoon scenario. Moreover, in case there is no platoon-members able to handle the computational task (e.g., platoon-members are overloaded), vehicles can always offload their tasks to the MA-RSU. Finally, the authors believe that platooning can provide powerful computation resources and achieve computing tasks in less time, as well as at higher efficiency.

### B. Platoon-Assisted vehicular Cloud for Vehicles

The proliferation of smart vehicles and their resource hungry applications impose a significant challenge on processing capabilities and service delivery of smart vehicles. To tackle this issue, we propose to leverage platoons to augment the computational capabilities of vehicular cloud. In fact, platoon-members with computation and communication capabilities can aggregate their computation resource and re-allocates them to satisfy the computation demands of individual vehicles. This is mainly because of the stability and predictability of platoons on the road. Therefore, platoons become abundant computation infrastructures, providing a great deal of computation resources.

In Fig. 3, vehicle discovers neighboring platoons within its communication range, and selects those in the same moving direction as candidates with sufficient stability (e.g., in terms of computation sufficiency and platoon string stability) on road. The platoon status including speed, location and moving direction, can be acquired through vehicular communication protocols. For example: in DSRC standard, the periodic beaconing messages can provide these state information. After selecting suitable platoon, vehicle uploads computation task for processing. Upon the completion of tasks execution, the selected platoon transmits back the result to the vehicle. In case that the vehicles are moving out of the V2V communications range of the platoon, the computation result can be sent back through nearest MA-RSU that the individual vehicle approaching.

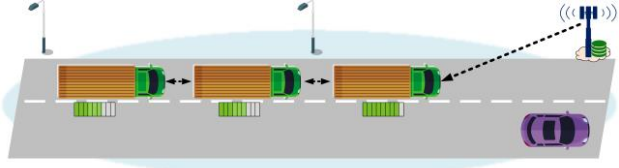

Fig. 4. Scenario of MEC offload its tasks fully or partially to platoons.

The integration of platoon cloud into vehicular cloud introduces some research challenges that needed to be addressed. The first challenge is the varying state of the vehicles during the time, i.e., they suddenly may increase/decrease their speed. For this purpose, designing efficient computation result retrieval strategy (i.e., considering other platoons as a relay or predictive mechanism) is necessary. Moreover, an efficient task scheduling and resource allocation scheme need to be investigated. For example: the platoon may be involved with heavy internal tasks. Therefore, processing additional tasks from other vehicles may lead to the overloaded resource which causing extra latency.

*C. Support for MEC*

In order to enhance the overall computing capacity, MEC may outsource part of its computing tasks to the platoons though RSU. This allows underutilized resources of vehicles in platoons provide support to execute complex applications and services in MEC. In this way, the vehicles in the platoon augment edge cloud computing eco-system by providing supplement cloud resources and effectively reduces computational load of MEC.

In Fig. 4, the MA-RSU discovers the potential candidates to assist the MEC and send the service requests through RSU. The discovery procedure is carried out through coordination between platoon leader and MEC. Therefore, MEC is aware of the platoon conditions (e.g., platoon stability and computing capability). Then, based on this information, tasks are assigned to the platoon(s) for processing. Finally, When the task computation is complete, the platoon sends the results back to MEC.

The exploitation of platoons to enhance MEC computation capacity and capability are followed by some research challenges. The main limitation is the delay caused by multi-hop communication, i.e., the mobile terminal offloads its task to MEC, and the MEC outsources the tasks to the platoon. As a result, offloading to the platoon through MA-RSU presents a significant challenge for delay-sensitive applications, such as virtual and augmented reality. Another important limitation of this scenario is the bounded coverage area of MEC, which restricts the capability of the platoon to provide computation assistance. In this case, the platoon needs to feedback the result before leaving the coverage area, which in turn make this procedure more complex. In contrast, the result of the computation tasks that can tolerate a certain period of latency, e.g., multimedia streaming and file backup, can be delayed up to the next connection of the platoon to the RSU.

In general, the limitations of this scenario make it infeasible to execute delay-sensitive applications. On the contrary, it can be a good candidate solution to provide service for delay-tolerant applications.

## V. CHALLENGES AND OPEN ISSUES

In this section, we highlight some of the most important challenges in realizing platoon-assisted vehicular cloud in VANET. These challenges are mainly communication, system sustainability, and security and privacy. We further align existing challenges for future research consideration. It is worth noting that platoon challenges can be classified into two generic groups: the first one covers the general platoon challenges and issues, such as platoon formation challenges, platoon control and stability, etc., which is outside the scope of this study (the interested readers are kindly referred to [31]–[34]); the second group is dedicated to the challenges raised in realizing platoon as a cloud; which is thoroughly discussed in this section.

*A. Communication*

Wireless communication is a critical factor in platooning. Usually, vehicles in platoon exchange messages that contain speed data, sensor readings, and current position to other vehicles or the infrastructure, and this is done using broadcasting or multicasting. Due to the dynamic nature of wireless communication, the network parameters, such as transmission delay and packet reception ratio are changing within a certain range. Thus, it is essential to develop robust and reliable network protocols for data dissemination in order to enable adaptive control of platoon-based cooperative driving system.

Another challenge is that the V2V and V2I are essentially designed to transmit safety messages with a small size of payloads. In contrast, the emerging application is/has typically big data size. As a result, the bandwidth of communication protocol in VANET become a bottleneck. Therefore, designing a high-bandwidth communication protocol is necessary.

*B. System Sustainability*

To ensure that platooning is sustainable, several aspects need to be taken into account. One aspect is to maximize the number of participants in platoons. Another aspect is to encourage more companies to engage with the development of platooning. Platooning could have implications for the heavily-used truck routes (e.g., supply chain network). Accelerating the development of ITS for transport network allows us to obtain the maximum benefits of platooning. Also, when designing a supply chain network, the platooning aspects need to be considered, such as the location of facilities that contributes to the additional benefits. Another interesting direction for future research is the development of incentive schemes to handle various restrictions on achieving sustainable platooning. The government could play an active role by providing incentive schemes to encourage the companies to participate in platoon development, for instance by subsidizing the technology or provide special cost reduction.

*C. Security and Privacy*

The security of platooning is another critical issue that needs to be addressed. Specifically, cooperative platoon-based driving system is more vulnerable to vicious attacks which may endanger the safety of road traffic. In such a distributed system, one vehicle may suffer potential attacks from other vehicles or infrastructures. The typical attacks include the malicious message, illegitimate commands, and the poisoning of map database locally stored on vehicles. To prevent malicious attacks, there is thus a need to set up an authentication system, in which sensible information from sensors need to be signed by private key [35]. Another way to deal with the security issue is to employ a misbehavior detection system [36].

## VI. CONCLUSION

This article has investigated a visionary concept of assisting vehicular cloud computing by incorporating platooning. The key contribution is to exploit the computation capacity of platoons, which is a group of vehicles with common interests, to augment the vehicular cloud for the nearby vehicles and passengers. The authors have also presented three potential scenarios and have discussed the exploitation of platoon resources and roadside infrastructure to facilitate new applications in emerging transportation system. Although, there exist a number of open research problems – in communication, security, privacy, and sustainability – that need to be solved, the vehicular platoon-assisted vehicular cloud is still expected to play a crucial role in the development and deployment of vehicular computation eco-systems in the short-term.

## ACKNOWLEDGMENT

This work has been performed with support of the H2020-MSCA-ITN-2015 project 5Gaura. This information reflects the consortia view, but the consortia are not liable for any use that may be made of any of the information contained therein.

## REFERENCES

[1] M. A. Javed, S. Zeadally, and E. Ben Hamida, "Data analytics for cooperative intelligent transport systems," *Veh. Commun.*, vol. 15, pp. 63–72, 2019.

[2] G. Karagiannis *et al.*, "Vehicular networking: a survey and tutorial on requirements, architectures, challenges, standards and solutions," *IEEE Commun. Surv. Tutorials*, vol. 13, no. 4, pp. 584–616, 2011.

[3] S. Hussain, Di Wu, S. Memon, and N. K. Bux, "Vehicular ad hoc network (vanet) connectivity analysis of a highway toll plaza," *Data*, vol. 4, no. 1, p. 28, 2019.

[4] A. Boukerche and R. E. de Grande, "Vehicular cloud computing: architectures, applications, and mobility," *Comput. Networks*, vol. 135, pp. 171–189, 2018.

[5] G. Qiao, S. Leng, K. Zhang, and Y. He, "Collaborative task offloading in vehicular edge multi-access networks," *IEEE Commun. Mag.*, vol. 56, no. 8, pp. 48–54, 2018.

[6] C. Krupitzer, M. Segata, M. Breitbach, S. El-Tawab, S. Tomforde, and C. Becker, "Towards infrastructure-aided self-organized hybrid platooning," in *2018 IEEE Global Conference on Internet of Things (GCIoT)*, 2018, pp. 1–6.

[7] DHL, "Platooning: one at the fore, all behind one," 2019. [Online]. Available: https://dhl-freight-connections.com/en/platooning-dhl-2/. [Accessed: 12-Jul-2019].

[8] N. E. Logistics, "Truck Platooning: Pros, Cons, and its Probable Future." [Online]. Available: https://nextexitlogistics.com/truck-platooning-pros-cons-and-its-probable-future. [Accessed: 12-Jul-2019].

[9] B. fuel, "Driving the future, advances in platooning technology." [Online]. Available: https://www.breakthroughfuel.com/blog/driving-the-future-advances-in-platooning-technology. [Accessed: 12-Jul-2019].

[10] Q. E. Ali, N. Ahmad, A. H. Malik, G. Ali, and R. Waheed, "Issues, challenges, and research opportunities in intelligent transport system for security and privacy," *Appl. Sci.*, vol. 8, pp. 1–24, 2018.

[11] L. Zhu, F. R. Yu, Y. Wang, B. Ning, and T. Tang, "Big data analytics in intelligent transportation systems: a survey," *IEEE Trans. Intell. Transp. Syst.*, vol. 20, no. 1, pp. 383–398, 2019.

[12] "Siemens eHighway -- Electrification of road freight transport," 2019. [Online]. Available: https://new.siemens.com/global/en/products/mobility/road-solutions/electromobility/ehighway.html. [Accessed: 12-Jul-2019].

[13] A. K. Bhoopalam, N. Agatz, and R. Zuidwijk, "Planning of truck platoons: A literature review and directions for future research," *Transp. Res. Part B Methodol.*, vol. 107, pp. 212–228, 2018.

[14] S. T. Rakkesh, A. R. Weerasinghe, and R. A. C. Ranasinghe, "An intelligent highway traffic model using cooperative vehicle platooning techniques," 2017.

[15] C. Shao, S. Leng, Y. Zhang, A. Vinel, and M. Jonsson, Eds., "Analysis of connectivity probability in platoon-based vehicular ad hoc networks," 2014.

[16] B. van Arem, C. J. G. van Driel, and R. Visser, "The impact of cooperative adaptive cruise control on traffic-flow characteristics," *IEEE Trans. Intell. Transp. Syst.*, vol. 7, no. 4, pp. 429–436, 2006.

[17] S. E. Shladover, "PATH at 20---history and major milestones," *IEEE Trans. Intell. Transp. Syst.*, vol. 8, no. 4, pp. 584–592, 2007.

[18] T. Robinson and E. Chan, "An introduction to the SARTRE platooning programme," in *17th World Congress on Intelligent Transportation System*, 2010, pp. 1–11.

[19] J. Larson, K.-Y. Liang, and K. H. Johansson, "A distributed framework for coordinated heavy-duty vehicle platooning," *IEEE Trans. Intell. Transp. Syst.*, vol. 16, no. 1, pp. 419–429, 2015.

[20] H2020-ART-2017, "Enabling safe multibrand platooning for Europe (ENSEMBLE)." [Online]. Available: www.platooningensemble.eu. [Accessed: 12-Jul-2019].

[21] "European Truck Platooning Challenge," 2016. [Online]. Available: https://eutruckplatooning.com. [Accessed: 12-Jul-2019].

[22] P. Kavathekar and Y. Chen, "Vehicle platooning: a brief survey and categorization," in *2011 ASME/IEEE International Conference on Mechatronic and Embedded Systems and Applications*, 2011, pp. 829–845.

[23] Arturo Davila and Mario Nombela, "Platooning - safe and eco-friendly mobility," in *SAE Technical Paper*, 2012.

[24] "3GPP TS 23.287 V0.2.0, Technical Specification Group Services and System Aspects; Architecture enhancements for 5G System (5GS) to support Vehicle-to-Everything (V2X) services. Release 16."

[25] E. Kulla, N. Jiang, E. Spaho, and N. Nishihara, *Complex, Intelligent, and Software Intensive Systems*, vol. 772. Cham: Springer International Publishing, 2019.

[26] S. Darbha, S. Konduri, and P. R. Pagilla, "Benefits of v2v communication for autonomous and connected vehicles," *IEEE Trans. Intell. Transp. Syst.*, vol. 20, no. 5, pp. 1954–1963, 2019.

[27] Hayder M. Amer, Christos Tsotskas, Matthew Hawes, Patrizia Franco, and Lyudmila Mihaylova, Eds., "A game theory approach for congestion control in vehicular ad hoc networks," 2017.

[28] "ETSI EN 302 663 V1.2.0, Intelligent Transport Systems (ITS); Access layer specification for Intelligent Transport Systems operating in the 5 GHz frequency band."

[29] H. Yao, C. Bai, D. Zeng, Q. Liang, and Y. Fan, "Migrate or not? exploring virtual machine migration in roadside cloudlet-based vehicular cloud," *Concurr. Comput. Pract. Exp.*, vol. 27, no. 18, pp. 5780–5792, 2015.

[30] X. Fan, T. Cui, C. Cao, Q. Chen, and K. S. Kwak, "Minimum-cost offloading for collaborative task execution of mec-assisted platooning," *Sensors (Basel).*, vol. 19, no. 4, 2019.

[31] D. Jia, K. Lu, J. Wang, X. Zhang, and X. Shen, "A survey on platoon-based vehicular cyber-physical systems," *IEEE Commun. Surv. Tutorials*, vol. 18, no. 1, pp. 263–284, 2016.

[32] E. Kulla, N. Jiang, E. Spaho, and N. Nishihara, "A survey on platooning techniques in VANETs," Springer, 2019, pp. 650–659.

[33] A. Soni and H. Hu, "Formation control for a fleet of autonomous ground vehicles: A Survey," *Robotics*, vol. 7, no. 4, p. 67, Nov. 2018.

[34] S. Darbha, S. Konduri, and P. R. Pagilla, "Benefits of v2v communication for autonomous and connected vehicles," *IEEE Trans. Intell. Transp. Syst.*, vol. 20, no. 5, pp. 1954–1963, May 2019.

[35] J. Blum and Azim Eskandarian, "The threat of intelligent collisions," *IT Prof.*, vol. 6, no. 1, pp. 24–29, 2004.

[36] J. Petit and S. E. Shladover, "Potential cyberattacks on automated vehicles," *IEEE Trans. Intell. Transp. Syst.*, vol. 16, no. 2, pp. 1–11, 2014.